\documentclass[a5paper, 11pt]{article}
\usepackage{graphicx}
\usepackage{fullpage}
\usepackage{fancyhdr}
\usepackage{enumerate}
\usepackage{mathtools}
\usepackage{amsmath}
\usepackage{txfonts}
\usepackage{gensymb}
\ProvideTextCommand{\DJ}{OT1}{\raisebox{0.25ex}{-}\kern-0.4em D}
\usepackage[titletoc]{appendix}

\usepackage[utf8]{inputenc}
\usepackage[swedish]{babel}
\usepackage{amssymb}
\usepackage[bottom=1.5cm,top=2.5cm,left=2.1cm,right=2.1cm]{geometry}
\usepackage{sectsty}
\usepackage[symbol]{footmisc}
\sectionfont{\large}

\addto\captionsswedish{}
\addto\captionsswedish{}
\addto\captionsswedish{}

\makeatletter
\def\@seccntformat#1{\csname the#1\endcsname.\quad}
\def\numberline#1{\hb@xt@\@tempdima{#1\if&#1&\else.\fi\hfil}}
\makeatother

\begin{document}
\pagestyle{fancy}
\renewcommand{\headrulewidth}{0pt}
\fancyhf{}
\cfoot{\text{\thepage}}

\thispagestyle{empty}
\clearpage
\pagenumbering{arabic} 

\Large
\noindent

\begin{center}
\textbf {Super-enhanced Nuclear Fusion\\ in Metal-like Systems and in Condensed Plasmas\\ of Supernova Progenitors}\footnote[1]{The author wishes to acknowledge the full support of research colleagues for his research in Japan and Sweden.}\footnote[2]{E-mail: hikegami@rcnp.osaka-u.ac.jp \\This article is an excerpt from the author’s book \textit{The Nature of the Chemonuclear Reaction} which is available by mail order from IkeTokyoTech@outlook.com.}
\end{center}
\large
\noindent
\begin{center}
\text{Hidetsugu Ikegami}
\end{center}
\footnotesize
\noindent
\begin{center}
\text{Professor Emeritus, Osaka University}
\text{Formerly Director, Research Center for Nuclear Physics, Osaka University}
\text{Honorary Doctorate, Uppsala University}
\end{center}

\newpage
\LARGE
\noindent
\setcounter{page}{1}
\begin{center}
\textsc{\textbf{1} Super-enhanced Nuclear Fusion in Metal-like Systems and\\ in Condensed Plasmas\\ of Supernova Progenitors}
\end{center}

%\makeatletter
%\g@addto@macro\thesection.
%\makeatother

\Large
\hyphenchar\font=-1
\section*{}
\normalsize

\noindent In the scheme of chemonuclear reaction, bulk of itinerant s-electrons revealing the thermodynamical liquid activity in metallic systems undergo contact interaction with atomic nuclei and therein nucleons, inducing contagiously \textit{thermodynamical liquid activity} among the bulk of nuclei under the irreversible action of Nature towards the chemical potential minimum resulting in the united spontaneous, irreversible atomic and nuclear reactions. The chemonuclear reaction, moreover, is enhanced with astronomical figures besides the enhancement of few particle processes (e.g., in the electron-screened pycnonuclear fusion). In the \textit{systems of metal-like hydride/deuteride and electron donor mixtures}, self-sustained H-H and D-D chemonuclear fusion may take place. In these systems, however, some unforeseen phenomena are induced, e.g., \textit{radiation-less fusion} and \textit{$\gamma$-ray missing positron annihilation}.

Furthermore in the systems, coherent $\mathrm{D_2}$-$\mathrm{D_2}$ and $\mathrm{D_3}$-$\mathrm{D_3}$ fusion take place with the enhancement factor over $10^{20}$$\sim10^{30}$ and $10^{30}$$\sim10^{46}$, respectively. Under the electron density correction on the chemical potential of reactants, the enhancement of chemonuclear fusion approaches to those of enhanced pycnonuclear fusion in the astrophysical condensed plasmas in stars such as whitedwarf supernova progenitors. It was found that \textit{chemonuclear fusion in the condensed matters revealing the thermodynamical liquid activity are essentially identical with the enhanced pycnonuclear fusion in the astrophysical condensed plasmas}. Both reactions are enhanced and reach astronom-\\ical figures in the presence of thermodynamical liquid activity. 

\hyphenchar\font=-1
\section*{1.1 Pycnonuclear and Chemonuclear Reactions in Metallic Liquids}
\noindent
Nuclear reactions have been known to be enormously enhanced in high density matters in stars [1]. The reactions are called pycnonuclear reactions after Cameron who coined the term from the Greek $\pi$ $\psi$ $\kappa$ $\nu$ $o$ $\sigma$, meaning “compact dense” [2]. In the condensed matters, electrons act to screen the Coulomb repulsion between the atomic nuclei and this screening effect becomes so remarkable that rates of reactions at low temeratures are almost independent of the temperature and mostly dependent on the density of matters (\textit{cf}. Eq (5) of Section 1.3).

In addition to this effect, the very cohesive effect manifested in solidification of dense liquids tends to enhance greatly the reaction rate [3, 4]. In metallic liquids, for example, in condensed plasmas in a white-dwarf progenitor of supernova, enhancement of the nuclear reaction rate by a factor of some tens figures of magnitude has been reported [5]. Substantial parts of the rate enhancement in the liquids are ascribable to this effect. While this enhancement is infeasible in gas plasmas like the solar interior, it is common to spontaneous reactions in liquids irrespective of kinds of the liquids and reactions as seen below.

In 2001, a possible occurrence of new enhanced pycnonuclear fusion was pointed out by the present author based on the microscopic considerations on the slow ion collision processes in the metallic Li-liquids [6]. In Uppsala, the author and R.Pettersson observed the enormous rate enhancement in the $\mathrm{^7Li(d,n)^8Be\rightarrow2\cdot}$$\mathrm{ ^4He}$ reaction by detecting $\alpha$-particles produced under some kV deuterium ion implan-\\tation on a Li-liquid target [7]. The observed enhancement was around $10^{10}$ which was predicted by taking into consideration the fusion proceeded through coupling with a spontaneous chemical, i.e., atomic reaction forming an intermediate united atom (quasi-Be atom) where twin oscillatory nuclei coexist at the center of common 1s-electron orbital. In the liquid the spontaneous reaction rate thereby the coupled nuclear reaction rate is enhanced by the factor $K\mathrm{= exp(-\Delta \textit{G}_r / \textit{k}_B \textit{T}})$ specified by the chemical potential (Gibbs energy) change $\Delta G_{\mathrm{r}}$ in the reaction. Here $\mathrm{\textit{k}_B}$ and $T$ denote the Boltzmann constant and the liquid temperature, respectively.

In the newly discovered scheme of enhanced pycnonuclear fusion in the Li-liquid, even the doubly intensified enhancement $K(\mathrm{Be}_2)=K^2(\mathrm{Be})$ with $\mathrm{\Delta \textit{G}_{r}(Be_2)= 2\Delta \textit{G}_r(Be)}$ was expected through forming intermediate quasi-$\mathrm{Be_2}$ molecules under the implantation of slow mol-\\ecular $\mathrm{D_2}$ ions [8]. This prediction was supported by a comparison experiment of atomic- and molecular-ion implantation on the same Li-liquid target by the author and T.Watanabe in Tokyo/Sakura [9]. They observed also an enormous rate enhancement close to $10^{20}$ as predicted [10] in the $\mathrm{^7Li(^7Li,2n)^{12}C}$ reaction [11].

These empirical results suggest that the enhanced pycnonuclear reactions take place also in the metallic Li-liquids and in condensed plasmas in the supernova progenitors. This conjecture was supported by thermodynamic data as seen in later. It would be of significance to investigate the enhancement mechanism of nuclear reactions in the metallic liquids such as metallic hydrogen and Li liquids from another point of view by taking into account the special characteristics of s-electrons that they interpenetrate the nuclei and therein come into contact interaction with the internal nucleons. This feature indicates the decisive influences of interpenetrating s-electrons on the nuclear kinetics. In fact, $\beta$-decays through the electron capture and electric monopole ($E0$) isomeric transitions through the internal s-electron conversion are the typical nuclear penetration effect of s-electrons [12].

Generally, bulk of valence atomic electrons reveal thermodynamic features and result in the rate enhancement of factors up to some tens figures of magnitude in the spontaneous chemical reactions in solutions. These features, namely the \textit{thermodynamical activity of liquid}, are likely to transfer contagiously into nuclear reactions with the collective dynamics of highly correlated itinerant s-electrons in metallic liquids or condensed plasmas under the irreversible actions of Nature towards the chemical potential minimum [13]. The mecha-\\nism of these enhanced reactions seemed to be the united spontaneous chemical and nuclear reactions in the metallic liquids or condensed plasmas.

At this point, it is necessary to remember the common features between the condensed plasmas in stars and metallic liquids that the number density of conduction or itinerant s-electrons range from $10^{22}$ to $10^{23}$ cm$^{-3}$ for the most light metals such as Li, Na and K. In these density regimes, effects arising from exchange and Coulomb coupling between the s-electrons are referred to as strongly coupled [14]. Due to the strong coupling or correlation effect within them, the itinerant s-electrons in these metallic liquids can be regarded as the Fermi liquids yielding the thermodynamical liquid activity.

As a matter of fact, in the molecular ion implantation experiments on the metallic Li liquids we observed the D-Li and $\mathrm{D_2}$-2Li fusion reactions enhanced by factors over 10 figures. Nowadays the reaction has been called '\textit{chemonuclear reaction}' [10]. Evidence of coherent $\mathrm{^8Be^*_2}$ formation in the $\mathrm{D_2}$-2Li reaction [15] and later corresponding heavy break-through phenomena was also obtained [9]. Here $\mathrm{^8Be^*}$ indicates a compound $\mathrm{^8Be}$ nucleus.

\hyphenchar\font=-1
\section*{1.2 Liquid Activity Revealed in Ionic- and Chemonuclear Reactions}
\noindent
Ionic reactions in aqueous solutions are typical spontaneous reactions and from the view point of two body reactions generally take place with enormously large rate enhancement e.g. as seen in the reaction [16],
\begin{align*}
\mathrm{Hg^{++} + S^{--} \leftrightarrow HgS(red)\,,} 
\end{align*}
\begin{equation}
K = \overrightarrow{k} / \overleftarrow{k} = 2 \times 10^{53} \: \: \mathrm{at \: 25^{\circ}C.}
\end{equation}
Here $K$, $\overrightarrow{k}$ and $\overleftarrow{k}$ denote the equilibrium constant, the forward- and backward-reaction rate, respectively. Equation (1) indicates that the forward spontaneous reaction rate $\overrightarrow{k}$ is enhanced.
\begin{equation}
\overrightarrow{k} = k_0 A \,\,.
\end{equation}
$A$ being the enhancement factor and $k_0$ being the intrinsic reaction rate between two reactant ions. While the backward reaction rate $\mathrm{\overleftarrow{k}}$ is reduced by the factor $A^{-1}$.\\

\noindent Recalling Eq.(1),
\begin{align*}
K = k_{0}A/(k_0/A) = A^2
\end{align*}
and thus
\begin{equation}
A=K^{1/2}.
\end{equation}

\hyphenchar\font=-1
Of interest is the fact that the same ionic reaction Eq. (1) takes place also in an alcohol solution with the same enhancement. Just in the both solvents, water and alcohol, $\mathrm{Hg^{++}}$ and $\mathrm{S^{--}}$  ions undergo the same ionic reaction Eq.(1) with the same enhancement. Even without solvent, the same ionic reaction takes place within fused salts. Generally the ionic reaction takes place with the large enhancement almost irrespective of kinds of solvents.

This empirical fact is well explained on the basis of the Widom’s general concept ''\textit{thermo-
dynamical activity of liquids}'' [13], which is common bulk/collective features of spontaneous reactions caused by the thermodynamic force in the liquids dissolving reactant particles (see Appendix 1). Here macroscopically distinct parts of the liquids surrounding the reactant particles are correlated and a long-range coherence appears. In general, the force is specified by the chemical potential (Gibbs energy) change $\Delta \textit{G}_r$ in the reactions. This general thermodynamic relation is strictly independent of kinds of reactant particles and the nature of microscopic interparticle interactions [17].

The mechanism of enhanced pycnonuclear reactions in metallic liquids as in condensed plasmas is in fact common to ionic reactions in liquids. It is natural that the observed chemonuclear reaction is the enhanced pycnonuclear reaction in the metallic Li-liquids.

We had, however, limited ourselves to the basic research, that is, ion implantation experiments on the metallic Li-liquid surfaces in order to investigate the reaction enhancement mechanism so far. This way of doing caused an instability of reaction rates peculiar to the chemically very reactive metallic Li-liquid surfaces. For instance, at a vacuum of $10^{-7}$Torr, it was hard to keep a clean Li surface within 1 sec during the slow D$_2$ ion implantation without dissociation of the $\mathrm{D_2}$ ions on the surface except for simultaneous slag film sputtering clean up treatment.

These empirical results of a series of experiments indicated the necessity of the introduction of a stabilizer of liquid Li metals such as Ni or Pd nanocrystal powders against the deterioration of chemically active liquid Li metals. The introduction of Ni or Pd nanocrystal powders combined with hydrogen opened the second phase of chemo-\\nuclear reaction research, because in combined systems of metal-like hydrides and Li, generally, a donor of itinerant electrons provides the possibility of self-sustained super-enhanced hydrogen-hydrogen nuclear fusion and successive He burning chemonuclear reactions.
\hyphenchar\font=-1
\section*{1.3 Hydrogen-Hydrogen Pycnonuclear Fusion Rates}
\noindent
We begin with the two body nuclear fusion between hydrogen (H/D) ions via a H-H bond or D-D bond collapse in the metal hydrogen systems. As detailed in the following section, the hydrogen ions are strongly screened by valence electrons and nearly localized itinerant electrons in hybridized states in the metal-like hydrides [18]. This effect is specified by the short-range screening length $D_{\mathrm{s}}$. Within the range, low energy fusion reactions are most effectively enhanced as seen below.

The ions are confined in respective bond spaces with a number density,
\begin{equation}
n_{\mathrm{i}} = (4\pi /3)^{-1} a_{\mathrm{i}} ^{-3},
\end{equation}
where $a_{\mathrm{i}}$ denotes the Wigner-Seitz radius of the ions [19]. 

The screened nuclear fusion rate per number density is given by,
\begin{equation}
R_{\mathrm{s}} = \lambda _{\mathrm{HH'}} n_{\mathrm{H}} n_{\mathrm{H'}} = \frac{2S(0)r_{\mathrm{HH'}}^* }{(1 + \delta _{\mathrm{HH'}}) \hbar} \cdot \sqrt{\frac{D_{\mathrm{s}}}{r_{\mathrm{HH'}}^*}} \mathrm{exp} \left[ -\pi \sqrt{\frac{D_{\mathrm{s}}}{r_{\mathrm{HH'}}^*}} \right] \cdot n_{\mathrm{H}} n_{\mathrm{H'}}\: ,
\end{equation}
between the hydrogen ions at low temperature [1]. Here $\lambda_{\mathrm{HH'}}$ denotes the fusion constant, $\delta_{\mathrm{HH'}} = 1$ for the same kind of hydrogen $\mathrm{H'=H}$ and 0 for a different kind of hydrogen $\mathrm{H'\neq H}$. The factor $S(0)$ refers to the reaction cross-section factor. The nuclear Bohr radius $r_{\mathrm{HH'}}^*$ is represented by the electron mass $m_{\mathrm{e}} = 0.511 \mathrm{MeV}/c^2$, the average nucleon mass $m\mathrm{_N = 931.5\,MeV}/c^2$, reduced ion mass $\mu _{\mathrm{HH'}}$ and the Bohr radius $a_{\mathrm{B}}$,
\begin{equation}
r_{\mathrm{HH'}}^* = \frac{m_{\mathrm{e}}}{2\mu_{\mathrm{HH'}} } \cdot a_{\mathrm{B}} = 1.45\times 10^{-14} \frac{A_{\mathrm{H}} + A_{\mathrm{H'}}}{A_{\mathrm{H}} A_{\mathrm{H'}} } \: \mathrm{(m)} \: ,
\end{equation}
provided,
\begin{align*}
\mu_{\mathrm{HH'}} = \frac{A_{\mathrm{H}} A_{\mathrm{H'}} m_{\mathrm{N}}}{A_{\mathrm{H}} + A_{\mathrm{H'}}} \: ,
\end{align*}
where $A_{\mathrm{H}}$ and $A_{\mathrm{H'}}$ denote the mass numbers of H- and H'- ions. 

\vspace{0.25cm}
\begin{center}
\small
\noindent Table 1.1 \, Cross-section factors $S(0)$ and Q-values
 \\of hydrogen burning fusion reactions [20, 21].
\vspace{0.25 cm}
\small
\begin{tabular}{| c | c | c |}
\multicolumn{3}{ p{7cm} }{} \\ \hline
 &  $S(0)$ $\mathrm{(MeV\cdot b)}$  & $Q$ (MeV) \\ \hline
$\mathrm{H(p, e^+ \nu_e)D}$ & $3.4\times 10^{-25}$ & $1.4$ \\ \hline
$\mathrm{D(p,\gamma )^3He}$ & $2.5\times10^{-7}$ & $5.5$ \\ \hline
$\mathrm{D(d,p)T}$ & $0.053$ & $4.0$ \\ \hline
$\mathrm{D(d,n)^3He}$ & $0.050$ & $3.3$ \\  \hline
$\mathrm{D(d,\gamma)^4He}$ & $\sim 10^{-7}$ & $23.8$ \\ \hline
\end{tabular}
\end{center}

The hydrogen burning fusion reactions, $S(0)$ and $Q$-values are tabulated in Table 1.1 [20, 21]. For these reactions, Eq.(5) is represented as
\begin{equation}
\mathrm{H(p,e^+ \nu_{\mathrm{e}})D} \: : \; R_{\mathrm{s}} = 1.5\times 10^{-45} n_{\mathrm{H}}^2 \sqrt{\frac{D_{\mathrm{s}}}{2.9\times 10^{-14}}} \mathrm{exp} \left[ -\pi \sqrt{\frac{D_{\mathrm{s}}}{2.9\times 10^{-14}}} \right] \: ,
\end{equation}
\begin{equation}
\mathrm{D(p, \gamma )^3He} \: : \; R_{\mathrm{s}} = 1.2\times 10^{-25} n_{\mathrm{H}} n_{\mathrm{D}} \sqrt{\frac{D_{\mathrm{s}}}{2.3\times 10^{-14}}} \mathrm{exp} \left[ -\pi \sqrt{\frac{D_{\mathrm{s}}}{2.3\times 10^{-14}}} \right] \: ,
\end{equation}

\begin{equation}
\mathrm{D(d, p)T} \: : \; R_{\mathrm{s}} = 1.2\times 10^{-22} n_{\mathrm{D}}^2 \sqrt{\frac{D_{\mathrm{s}}}{1.5\times 10^{-14}}} \mathrm{exp} \left[ -\pi \sqrt{\frac{D_{\mathrm{s}}}{1.5\times 10^{-14}}} \right] \: ,
\end{equation}

\begin{equation}
\mathrm{D(d, n)^3He} \: : \; R_{\mathrm{s}} = 1.1\times 10^{-22} n_{\mathrm{D}}^2 \sqrt{\frac{D_{\mathrm{s}}}{1.5\times 10^{-14}}} \mathrm{exp} \left[ -\pi \sqrt{\frac{D_{\mathrm{s}}}{1.5\times 10^{-14}}} \right] \: ,
\end{equation}

\begin{equation}
\mathrm{D(d, \gamma)^4He} \: : \; R_{\mathrm{s}} \approx 10^{-28} n_{\mathrm{D}}^2 \sqrt{\frac{D_{\mathrm{s}}}{1.5\times 10^{-14}}} \mathrm{exp} \left[ -\pi \sqrt{\frac{D_{\mathrm{s}}}{1.5\times 10^{-14}}} \right] \: .
\end{equation}

\noindent where $R_{\mathrm{s}}$, $n$ and $D_{\mathrm{s}}$ are given in the units of $\mathrm{s^{-1} m^{-3}}$, $\mathrm{m^{-3}}$ and m respectively.

\hyphenchar\font=-1
\section*{1.4 Metallic Hydrogen in Metal-like Hydrides}
\noindent
\textit{In the early 1970s, the possible formation of metallic hydrogen corresponding to the aforementioned hybridized states in some transition metals was pointed out based on the systematic investigation of alloys} [22]. The forming of transition metal hydride takes place in two steps of chemisorption: first, hydrogen molecules dissociate at the surface of adsorbent Pd or Ni nano-clusters and are transformed into metallic hydrogen; second, the metallic hydrogen is alloyed. Corresponding to this investigation, transition metal hydrides revealing metallic features will be called \textit{metal-like hydrides} in this book.

This explanation becomes relevant because the electronegativity (Pauling Scale) of hydrogen $\chi$=2.20 which indicates the power of an atom to attract electrons to itself is close to those of Ni ($\chi$=1.98) and Pd ($\chi$=2.20) [16] resulting the good atomic size matching for the alloy formation. On the contrary the electronegativity of Li ($\chi$=0.98) and Ca ($\chi$=1.00) differ from those of Ni and Pd. As a result, for transferring the Li and Ca s-electrons to the Pd and Ni cells to form alloy the size of Li and Ca atoms becomes considerably smaller than the cells so that this mismatching at boundaries between Li/Ca and Ni/Pd metals become less relevant. 

Instead, Li and Ca supply the itinerant electrons in the Ni and Pd metals, which induce the lowering of melting point of adsorbed metallic hydrogen leading to the enhanced cold nuclear fusion of metallic hydrogen liquid. The same benefit is expected for Na, K, Cs, Sr and Ba. In the presence of little itinerant electrons, the metallic hydrogen forms a quantum solid rather than a semi classical liquid in the metal hydrides (e.g., deuterons perform zero-point oscillation around their lattice sites), and their fusion rates are determined as in Eqs.(9-11) from the contact probabilities between adjacent nuclei. However, the rate is extremely faint as seen later. 

For the manifestation of enhancement arising from the thermody-\\namical liquid activity, at least one of the reacting nuclear species must be in a liquid state. A melting temperature ${T_m}$ derived from a criterion of Lindemann type [14] for hydrogen D may be written in the form, under Thomas-Fermi approximation,
\begin{align*}
3k_{\mathrm{B}}T< \hbar^2 \lbrack \frac{(3\pi n_{\mathrm{e}})^ \frac{2}{3}}{2m}\rbrack ,
\end{align*}
\begin{equation}
T_{\mathrm{m}}= (\frac{e^2}{180a_{\mathrm{D}}k_{\mathrm{B}}})\{1+(\frac{20}{3\pi})(\frac{\alpha}{\lambdabar_{\mathrm{e}}})(\frac{n_{\mathrm{e}}}{12n_{\mathrm{D}}^2})^\frac{1}{3}\}^{-1} .
\end{equation}

${a_D}$ : Wigner-Seitz radius of the deuteriums,

${\lambdabar_e}$ : Compton wavelength of the electrons,

${n_e}$ : number density of the electrons,

$m$ : electron mass 

$\alpha$ : fine structure constant. 

\hyphenchar\font=-1
\section*{1.5 Metallic Hydrogen Liquids in Metal-like Hydride- \\ElectronDonor Mixtures $-$ the Cold Nuclear Fusion Puzzle}
\noindent
In Eq.(12), the first bracket indicates the intrinsic melting temperature of metallic deuterium in the metal-like hydrides associated with little density itinerant electrons. For instance, in PdD of $\mathrm{n_D}$ $\approx$ $6.3\times 10^{28}m^{-3}$ referring Table 1.2, the intrinsic melting temperatures is ${T_m}$ = $\frac{e^2}{180a_{\mathrm{D}}k_{\mathrm{B}}}$ $\approx$ 300K, which implies that the manifestation of liquid activity of metallic deuterium is critical at room temperature. However, in the metal-like hydrides built in itinerant electrons of ${{n_e}}$ $\approx$ $10^{26} m^{-3}$, the melting temperature of metallic deuterium reduces to about 200K and the metallic deuterium reveals the thermodynamical liquid activity at room temperature, which results in the fusion rate enhancement of factor over $10^{30}$ as described in Sections 1.8-1.10. Above argument interprets well the long-pending puzzle regarding the unrepeatable cold nuclear fusion experiments. 
\vspace{0.25cm}
\begin{center}
\small
\noindent Table 1.2 \,Parameters in Eqs. (7) - (11) for the screened low temperature nuclear fusion rate and power released in metal hydrogen systems [14].
\vspace{0.25 cm}
\small
\begin{tabular}{| c | c | c | c |}
\multicolumn{4}{ p{9.6cm} }{} \\ \hline
 & $\mathrm{TiH_2/TiD_2}$ & NiH/NiD & PdH/PdD \\ \hline
$n\mathrm{_H / }n\mathrm{_D (m^{-3})}$ & $9.4\times 10^{28}$ & $7.3\times 10^{28}$ & $6.3\times 10^{28}$ \\ \hline
$D\mathrm{_s (10^{-11}m)}$ & 2.8 & 2.2 & 1.9 \\ \hline
$R_s\mathrm{(s^{-1} \cdot m^{-3}), H-H}$ & $1.7\times 10^{-28}$ & $5.8\times 10^{-24}$ & $1.8\times 10^{-21}$ \\ \hline
$R_s\mathrm{(s^{-1} \cdot m^{-3}), D-D}$ & $1.0\times 10^{-21}$ & $2.6\times 10^{-15}$ & $8.9\times 10^{-12}$ \\ \hline
$P\mathrm{(W \cdot m^{-3}), H-H}$ & $1.5 \times 10^{-41}$ & $5.2\times 10^{-37}$ & $1.7\times 10^{-34}$ \\ \hline
$P\mathrm{(W \cdot m^{-3}), D-D}$ & $5.9\times 10^{-34}$ & $1.5\times 10^{-27}$ & $5.2\times 10^{-24}$ \\ \hline
\end{tabular}
\end{center}

In the experiments such as those of electrochemical method, a nano-structured Pd cathode and an alkali-${\mathrm{D_2}}$O solution e.g. Cs${\mathrm{NO_3}}-{\mathrm{D_2}}$O solution is used. At the surface of Pd cathode, ${\mathrm{D_2}}$O is reduced producing deuterium which undergoes chemisorption forming PdD. Meanwhile Cs${\mathrm{NO_3}}$ is also reduced but most of the produced Cs atoms immediately react with the solution forming CsOD. But very faintish fractions of Cs atoms are taken into Pd cathode and perform a part of electron donor. However, usually produced itinerant electrons are slender and resulted their action is very critical as argued above for the lowering the melting temperature of adsorbed metallic deuterium. Currently used LiOD-${\mathrm{D_2}}$O solution electrochemical method is also the case. 

The most useful treatment to solve this problem is the introduction of metal-like hydrides - electron donor mixtures. One of the simplest way to prepare the mixture may be a combination of Ni nanocrystal powder and LiD or its complex compound e.g. LiAl${\mathrm{D_4}}$ or LiMg${\mathrm{D_3}}$ in an atmosphere of ${\mathrm{D_2}}$ gas. 

\hyphenchar\font=-1
\section*{1.6 Hydrogen-Hydrogen Chemonuclear Fusion via Hydrogen Bond Collapse}
\noindent
In the fusion system of metal-like hydride-electron donor mixture, dimensionless de Broglie wave length $\Lambda _{\mathrm{i}}$ of hydrogen atoms with mass number $A\mathrm{_i}$ is very small.
\begin{equation}
\Lambda_{\mathrm{i}} = \frac{\hbar}{a_{\mathrm{H}}} \left( \frac{2\pi}{A_{\mathrm{i}} m_{\mathrm{N}} k_{\mathrm{B}} T} \right)^{\frac{1}{2}} < 1 \: .
\end{equation}
where $a_{\mathrm{H}}$ denotes the Wigner-Seitz radius in Eq.(4). 

The wave mechanical effects are therefore negligible on the atoms. Furthermore, hydrogen pairs reacting within their bonds do not disturb so much surrounding spectator atoms in the molecules because their interaction can be treated as that between two screened particles. These features confirm the validity of rate enhancement evaluation in the scheme of semi-classical dynamics as far as concerning the nuclear fusion under the condition of hydrogen ions. Their dynamics may be describable as those of electron screened ion cores immersed in the sea of itinerant electrons. 

Within the bonds, the hydrogen atoms collide with each other e.g. at $T$ = 773 K with the frequency,
\begin{equation}
\nu_{\mathrm{H-H}} = k_{\mathrm{B}} T / \hbar = 1\times 10^{14} \: \mathrm{s^{-1} }\: ,
\end{equation}
which is much smaller than the gyration frequency $\nu_{\mathrm{s}}$ of 1s-orbital electrons of hydrogen atoms,
\begin{equation}
\nu_{\mathrm{s}} = \upsilon_{\mathrm{B}} / 2\pi a_{\mathrm{B}} = \alpha c / 2\pi a_{\mathrm{B}} = 7\times 10^{15} \: \mathrm{s^{-1}} \: ,
\end{equation}
where $\upsilon_{\mathrm{B}}=\alpha c$ denotes the Bohr speed, c being the speed of light. This means that the orbital electrons adjust their electronic states continuously and smoothly to the nuclear collision processes. Thereby nuclear fusion reactions take place via collapse of the hydrogen bonds forming intermediate united atoms (quasi-He atoms) in which pairs of oscillatory hydrogen nuclei coexist at centers of common 1s-orbitals. Such united atoms are commonly known through x-ray spectroscopic measurements in ion-ion colliding experiments.

During the collapse, the volume of two screened hydrogen atoms shrinks by one order of magnitude in association with the chemical potential change $\Delta G_{\mathrm{r}}$ [10]. Normally the most quasi-He atoms do not necessarily proceed to the nuclear fusion and decay into pairs of hydrogen atoms as before. In the chemonuclear fusion system, however, the hydrogen bonds are metallic and liquified in the sea of itinerant s-electrons, which reveal the thermodynamic activity of liquid and sustain the shrinking and thereby prolong live times of the quasi-He atoms by the factor $A= K^{1/2} = \mathrm{exp}\left[ -\Delta G_{\mathrm{r}} / 2k_{\mathrm{0}} T \right]$ as seen in Eq.(3). This results in the enormously enhanced rate $KR_s$ of chemonuclear fusion caused by the zero-point oscillation, recalling Eqs.(7-11) as will be explained in detail in Sections 1.8-1.10.

The above result is based on the general thermodynamic relation that the fractional change of reaction rate is exactly proportional to the entropy change in the universe [17].

The above arguments are based on the chemical potential (Gibbs free energy $\Delta G_r$) and hereby the chemical potential change $\Delta G_r$ in the reaction defined at the normal electron density on earth $n_e=0.03$ atomic unit $= 0.03(\frac{4\pi}{3})^{-1}a_\mathrm{B}^{-3}$. However, the itinerant s-electron density in such metal-like hydride and electron donor as Li mixtures is very high depending on the kinds of electron donor and its mixing amount, for instance, $n_e = 0.085$ a.u.

Since the chemical potential is proportional to $n_e^{2/3}$, the chemical potential and hereby the chemical potential change in the mixtures should be replaced as

\begin{align*}
\Delta G_r \mathrm{(normal}\,n_{e^-}) \longrightarrow \Delta G_r(n_{e^-}=0.085\, \mathrm{a.u.})
\end{align*}
\begin{align*}
= (0.085/0.03)^{2/3}\Delta G_r(\mathrm{normal}\,n_{e^-})=2.0\Delta G_r \, .
\end{align*}
\noindent
It is thus reasonable to employ the equilibrium constant $K$ as the enhancement $A$ in the dense electron donor mixtures.
\begin{equation}
A=\mathrm{exp}\,[-2.0\Delta G_r/2k_\mathrm{B}T]=\mathrm{exp}\,[-\Delta G_r/k_\mathrm{B}T]=K\,.
\end{equation}
Eq.(16) denotes the enhancement of chemonuclear reaction in the metal-like hydride-electron donor mixtures in this book, where $\Delta G_r$ being $\Delta G_r$ (normal $n_{e^-}$) unless otherwise specified.

\hyphenchar\font=-1
\section*{1.7 Elementary Statistical Consideration of Chemonuclear Fusion}
\noindent
Though the rate enhancement in the nuclear fusion $(\Delta G_\mathrm{r} < 0)$ has been derived exactly by astroplasma physicists based on the quantum statistical mechanics [1,3,4] it would still be useful to reconsider the enhancement mechanism from the elementary viewpoint.

As a postdoctoral fellow of Rutherford's laboratory, Bohr, who developed the concepts of \textit{nucleus} and \textit{Rutherford scattering} [23], explained that the latter was not simply collision between nuclei but was collision between whole atoms at low energies, typically several keV/amu on the basis of his atomic model [24]. This phenomenon is called nuclear stopping of ions/atoms moving in a condensed matter nowadays, because the nuclei of struck ions/atoms acquire significant amounts of kinetic energy in the collision as seen in the sputtering phenomena of ions [25]. This implies that collisions of nuclei are fully linked to atomic collisions which are strictly governed by the statistical thermodynamics as seen in chemical reactions.

This feature of nuclear stopping of ions/atoms keeps a hidden potential leading to the new concept 'chemonuclear reactions', where nuclear reactions take place united with the atomic processes and thus are controllable through the synchronous formation of atomic- and nuclear- intermediate complexes under some proper chemical and physical conditions [6-12].

The above mentioned state of affairs would be visualised if one considers the D-D atomic collision in a condensed matter from the microscopic viewpoint. The collision is essentially the Rutherford scattering between two deuterons dressed with atomic electrons. They are directed to the nuclear collision at their screening distance $\mathit{D_s}$ e.g. $\mathit{D_s=}$ 19(pm) in a Pd nano-cluster powder (\textit{cf.} Table 1.2). This means that the colliding D-atoms most likely form a united atom (quasi-He atom) where twin deuterons are co-existing and confined at the center of common quasi-He K-shell electron orbital of radius 26(pm). In the quasi-He atom twin deuterons are colliding with each other with the zero-point oscillation energy towards the pycnonuclear fusion through the tunneling effect.

The united atom (quasi-He atom) corresponds to the intermediate complex which decays into the final product in the reaction and thus we may apply the rate equation of chemical reaction via intermediate complex formation also to the united atom formation. Here, well known Arrhenius rate equation is represented in the form recalling Eq.(16),
\begin{align*}
\overrightarrow{k} = k_0 \,\mathrm{exp}[-\Delta \textit{G}_r / \textit{k}_BT] = k_0K \tag{16'}
\end{align*}
in the scheme of Gibbs statistical thermodynamics. The chemical potential change $\Delta G_r$ in the reaction corresponds to the activation energy $E_a$ in the Arrhenius equation and $\Delta G_r > 0$ in the reversible reaction while $\Delta G_r < 0$ in the spontaneous reaction revealing the reaction enhancement $K > 1$.

\hyphenchar\font=-1
\section*{1.8 Enhancement Factor Evaluation of Chemonuclear Fusion Rate}
\noindent
For the fusion reaction producing a compound He nucleus via a quasi-\\He atom formation,
\begin{equation}
\mathrm{D + D \longrightarrow quasi\mbox{-}He \longrightarrow He} \: ,
\end{equation}
the chemical potential change $\Delta G_{\mathrm{r}}$ is evaluated by,
\begin{equation}
-\Delta G_{\mathrm{r}} (\mathrm{quasi\mbox{-}He} ) = -\Delta G_{\mathrm{r}} (\mathrm{He}) = \phi^* (\mathrm{He}) - 2 \phi^* (\mathrm{D}) .
\end{equation}
The change of chemical potential $\phi^*$ of impurity atoms in bulk metals [22] may be approximated by the change of immersion energy $\Delta E^{hom}$ of atoms embedded in an electron gas/liquid. The immersion energy has been evaluated as a function of the electron density (in the atomic 
unit : $\frac{3}{4\pi{{a_B}}^3}=1.6\times10^{30} \mathrm{m}^{-3}$). 

At lower densities such as ${{n_e}} < 0.01\,a.u.$, $\Delta E^{hom}$ becomes negative with a minimum point for the atoms with stable free negative ions such as H$^-$ ions [26]. 
Accordingly one may evaluate the plausible value $\phi^*$(He) using the empirical formula Eq.(19) derived from the systematics of alloy data on $\phi^*$, Li: 2.85(eV), Be: 4.20 (eV), B: 4.80 (eV), C: 6.23 (eV), N: 7.00 (eV), [32]. 
\begin{equation}
\phi^* = 1.25(Z-1)+0.03 \textrm{(eV)},
\end{equation}
\begin{equation}
\phi^*\textrm{(He)} = 1.28 \textrm{(eV)}.
\end{equation}
In the same way,
\begin{equation}
\phi^*\textrm{(D)} = \phi^*\textrm{(H)} = 0.03 \textrm{(eV)}.
\end{equation}
Then we have,
\begin{equation}
-\Delta G_{\mathrm{r}}\textrm{(He)} = 1.22 \textrm{(eV)}.
\end{equation}

Recall Eq.(16), we have the enhancement factor of chemonuclear D-D fusion via the collapse of D-D bond immersed in the fusion system as,
\begin{equation}
K(\mathrm{He}) = 2.9\times 10^{20} \: \: \mathrm{at} \: T = 300\mathrm{K} \: (k_{\mathrm{B}} T = 0.0259 \: \mathrm{eV}).
\end{equation}
The enhancement is reduced at higher temperatures, for instance, 
\begin{equation}
K(\mathrm{He}) = 2.4\times 10^{13} \: \: \mathrm{at} \:  T=460K .
\end{equation}

\noindent The above treatment may, however, result in underestimated chemical potential -- in the system of metal-like hydride-electron donor mixture and Ni hydride-Li mixture in particular, the interstitial site hydrogen atoms are metallic [22]. If we apply another empirical formula derived from systematics of alloy data to metallic elements Li and Be, instead of Eqs.(19-24), we get, respectively,
\vspace{-0.1cm}
\begin{align*}
\phi^* = 1.40(\mathrm{Z}-1) \,(\mathrm{eV}), \tag{19'}
\end{align*}
\begin{align*}
\phi^*(\mathrm{He}) = 1.40\,(\mathrm{eV}),  \tag{20'}
\end{align*}
\begin{align*}
\phi^*(\mathrm{D})=\phi^*(\mathrm{H})=0 \tag{21'}
\end{align*}
\begin{align*}
-\Delta G_\mathrm{r}\mathrm{(He)} = 1.40 \, (\mathrm{(eV)}, \tag{22'}
\end{align*}
\begin{align*}
K\mathrm{(He)} = 1.4\times10^{25}\,\mathrm{at\,T=300K} \tag{23'}
\end{align*}
\begin{align*}
K\mathrm{(He)} = 2.2\times10^{15}\,\mathrm{at\,T=460K} \tag{24'}
\end{align*}

\hyphenchar\font=-1
\section*{1.9 Evidence of Coherent Chemonuclear Fusion}
\noindent
In the scheme of chemonuclear fusion where deuterium ion clusters undergo the fusion, the rate enhancement may further be intensified by the coherent bond collapse [8]. As a matter of fact, we observed a coherent chemonuclear fusion of molecular $\mathrm{D_2}$ ions implanted with a slow speed $v_{\mathrm{i}} < \alpha Z_{\mathrm{Li}} c$ on a Li-liquid surface [9]. During the collision, pairs of atoms in the ions keep their correlation and induce coherent atomic/ionic collisions. These coherent collisions were already known in the 1970s, for instance, in the coherent sputtering phenomena of molecular ions [27].

It is worth noting that the observed rate enhancement of molecular $\mathrm{D_2}$ ion induced fusion was close to the square of rate enhancement of deuteron induced fusion in our experiment [9]. In the molecular $\mathrm{D_2}$ ion induced fusion, we observed the coherent production of twin $\alpha$-particle pairs which indicate the coherent decay of an intermediate compound nuclear pair $\mathrm{(^8 Be)_2}$ [15]. The author therefore remembers the same kind of phenomenon -- the coherent decay of positronium molecule $\mathrm{Ps_2}$ into twin degenerate annihilation photon pairs [28].

\hyphenchar\font=-1
\section*{1.10 Intensified Rate Enhancement of Coherent $\mathrm{D_n}-\mathrm{D_n}$ Chemonuclear Fusion}
\noindent
For the coherent $\mathrm{D_n}$-$\mathrm{D_n}$ chemonuclear fusion,
\begin{equation}
\mathrm{D_n + D_n} \longrightarrow (\mathrm{quasi\mbox{-} He})_n\longrightarrow (\mathrm{He})_n \: ,
\end{equation}
the chemical potential change $\Delta G_{\mathrm{r}} (\mathrm{He_n})$ is given by recalling Eqs. (22) and (22'),
\begin{equation}
\Delta G_{\mathrm{r}} (\mathrm{He_n}) = n\Delta G_{\mathrm{r}} (\mathrm{He}) = -1.22n \: (\mathrm{eV}) \sim -1.40n\, \mathrm{(eV)}.
\end{equation}
It may be said in this connection that He$_n$ molecules consisting of excited He atoms are known to be stable. The enhancement factors of chemonuclear fusion at the temperature $T=460\mathrm{K}$ via the $n$-fold, doubly -and trebly- coherent collapse of D-D bonds are, respectively,
\begin{equation}
K(\mathrm{He_n}) = K^n (\mathrm{He}) = (2.4\times 10^{13})^n \: ,
\end{equation}
\begin{equation}
K(\mathrm{He}_{2}) = 5.7\times 10^{26} \: , \: K(\mathrm{He}_{3}) = 1.4\times 10^{40}.
\end{equation}

In particular, in the system of Ni hydride-Li mixture, Eqs.(27) and (28) may be replaced by, respectively, recalling Eq.(24'),
\begin{align*}
K(\mathrm{He_n}) = K^n (\mathrm{He}) = (2.2\times 10^{15})^n \: , \tag{27'}
\end{align*}
\begin{align*}
K(\mathrm{He}_{2}) = 5.1\times 10^{30} \: , \: K(\mathrm{He}_{3}) = 1.2\times 10^{46}. \tag{28'}
\end{align*}

\noindent In the case of mixed H and D systems, however, a coherent collapse of two H-D bonds is unlikely because another coherent H-H or D-D collapse is more favorable. The $\mathrm{D(p,\gamma)^3He}$ reactions via the coherent H-D bond collapse thus are unrealized.

\hyphenchar\font=-1
\section*{1.11 D-D Chemonuclear Fusion through Coherent $\mathrm{H_n}$-$\mathrm{H_n}$ Chemonuclear Fusion}
\noindent
In the Li permeated transition metal/hydrogen systems with neither deuteride nor $\mathrm{D_2}$ gas, the D-D chemonuclear fusion still takes place through the coherent $\mathrm{H_2}$-$\mathrm{H_2}$ chemonuclear fusion. In the $\mathrm{H(p,e^+\nu_e)D}$ reactions via the H-H bonds’ collapse, D atoms/ions are produced at rest because their recoil energy in the positron and neutrino emission is below 0.1 eV. This implies that new D-D bonds are formed through the coherent $\mathrm{H_2}$-$\mathrm{H_2}$  chemonuclear fusion in the systems.

The D atoms/ions diffuse in the metal hydrogen systems, e.g., the diffusion coefficient reaches $D_{\mathrm{o}} = 4\times 10^{-8} \: \mathrm{m^2 /s}$ in the $\mathrm{Mg_2 NiD_4}$ crystal [29]. This implies that the D atoms produced prefer to cluster at the same sites in the crystal due to Bose-Einstein condensation, and this phenomenon leads to the coherent $\mathrm{D_n}$-$\mathrm{D_n}$ chemonuclear fusion.
Incidentally the author may remark from his experience that some ten-percent fraction of hydrogen ions produced in conventional PIG type ion sources are the molecular ions H$_2^+/$D$_2^+$ and H$_3^+/$D$_3^+.$
These features provide us with a new control scheme of energy released in the fusion through tuning precisely the fraction of $\mathrm{D_2}$ gas because the intrinsic H-H and D-D nuclear fusion rates differ by a factor of some billions in Ni- and Pd-hydrides, as seen in Table 1.2.

Table 1.2 lists those parameters pertinent to the electron screened nuclear fusion reactions [14]. The values adopted for NiD have been estimated from those for Ti$\mathrm{D_2}$ and PdD based on the data which indicate that both the heat of hydrogen dissolution and the lattice parameter of Ni are closer to those of Pd than Ti [29,30].

Note that the $\mathrm{D(p,\gamma)^3 He}$ reaction has been disregarded because any significant contribution caused by the coherent bond collapse is unlikely in the H-D chemonuclear fusion as argued in Section 1.10.

In the cases of Pd and Ni (Table 1.2), the powers released in the D-D fusion are bigger than those in the respective H-H fusion by a factor over one billion. This implies that in a system of Li permeated metal and pure $\mathrm{H_2}$ the D-D chemonuclear fusion still takes place. The rate of chemonuclear fusion shows clearly that almost all energy released in the pure hydrogen system is due to the D-D fusion in succession of the coherent H-H fusion and the D-D fusion caused by D$_2$ gas which is found by a fraction of about $\frac{1}{6000}$ of ordinary hydrogen gas. Most of the experiments in which cold fusion was observed successfully support this explanation.

\hyphenchar\font=-1
\section*{1.12 System of Metal-like Hydride-Electron Donor Mix-\\tures}
\noindent
The thermodynamic activity of liquids, revealed in the condensed plasmas in stars such as supernova progenitors, may be reproduced in the combined system of electron liquids and metallike hydrides. It has long been known that some transition metal hydrides are metal-like and therein hydrogen atoms are partly in the metallic state [22,31,32]. About some tenths percent fraction of these hydrogen atoms are likely itinerant and reveal the thermodynamic liquid activity in the presence of dense itinerant electrons as argued in Sections 1.4-1.5.

For example, we could introduce hydrogen adsorbing Ni or Pd nano-cluster powders or nano-structured foils combined with Li or Na metals and/or alkaline earth oxides CaO or SrO as the donors of dense electrons. Especially Li metal will be able to play important roles as the source of itinerant ions and electrons corresponding to the metallic hydrogen liquids in the white-dwarf supernova progenitors. The Li metal can be replaced by Li hydride LiH(D) or its complex (e.g., LiAlH$_4$(D$_4$)) because these hydride molecules dissociate at the surface of nano-clusters and liberated hydrogen atoms are adsorbed in them. This is due to the fact that Li hydride bond strength is smaller than the sum of bond strength of NiH(D) and NiLi. With this prescrip-\\tion [34], Bolognia and Uppsala collaboration group increased the amount of Li and $\mathrm{D_2}$ and improved their experimental results [35,36].

At an appropriate fraction of itinerant ions and electrons, their correlation among the reactant atoms is so strong that their kinematics is no longer those of isolated two-body interaction and reveal the metallic liquid activity as argued in Sections 1.6-1.9.

\hyphenchar\font=-1
\section*{1.13 Radiation-less Coherent $\mathrm{D_2}$-$\mathrm{D_2}$ and $\mathrm{D_3}$-$\mathrm{D_3}$ Chemonuclear Fusion}
\noindent
In the coherent $\mathrm{D_2}$-$\mathrm{D_2}$ and $\mathrm{D_3}$-$\mathrm{D_3}$ chemonuclear fusion, the respective rates forming $\mathrm{He_2}$ and $\mathrm{He_3}$ complexes are, recalling Eq.(11),
\begin{equation}
R(\textrm{D}_2-\textrm{D}_2) = 3.3 \times 10^{-28} n^2_{D_2} \cdot  fK^2
\end{equation}
\begin{equation}
R(\textrm{D}_3-\textrm{D}_3) = 3.3 \times 10^{-28} n^2_{D_3} \cdot  fK^3
\end{equation}
Here, $f$ denotes the penetration factor given as
\begin{equation}
f = \sqrt{\frac{D_S}{1.5\times10^{-14}}} \: \mathrm{exp} \:  \Bigg[ -\pi \sqrt{\frac{D_S}{1.5\times10^{-14}}\Bigg]},
\end{equation}
and $K$ = $2.4\times10^{13}$ at $T$=460K in Eq.(24) or $K$ = $2.2\times10^{15}$ in Eq.(24') .  
The remarkable features of this coherent fusion are the radiationless decay of intermediate fusion complexes quasi-$\mathrm{He_2}$ and quasi-$\mathrm{He_3}$ which undergo respective coherent decays into double and triple He ions with kinetic energy of 23.8 MeV/He ion without any recoil effect. 
All these He ions induce successive reactions resulting in the element synthesis as seen in the astrophysical element synthesis processes, which open the very powerful way of radioactive waste vanishing as seen in Chapter 2. Also noted are the mass synthesis of noble elements as seen in Chapter 4.

\hyphenchar\font=-1
\section*{1.14 Dynamics in Ni Hydride-Li Mixtures and $\gamma$-ray missing Positron Annihilation}
\noindent
A fusion test system of $\mathrm{Mg_2 Ni H_4}$/$\mathrm{Mg_2 Ni D_4}$ is raised as an example. For the system we may apply the treatment for the hydride NiH(D) presented in Section 1.12. The crystal structure of $\mathrm{Mg_2 NiD_4}$ is shown in Fig 1.1. In the crystal, hydrogen (H/D) ions cluster up to six at octahedral (O) sites and indicate the possibility of trebly coherent $\mathrm{D_3}$-$\mathrm{D_3}$ chemonuclear fusion.

\begin{figure}[h!]
\centering
\includegraphics[scale=0.2]{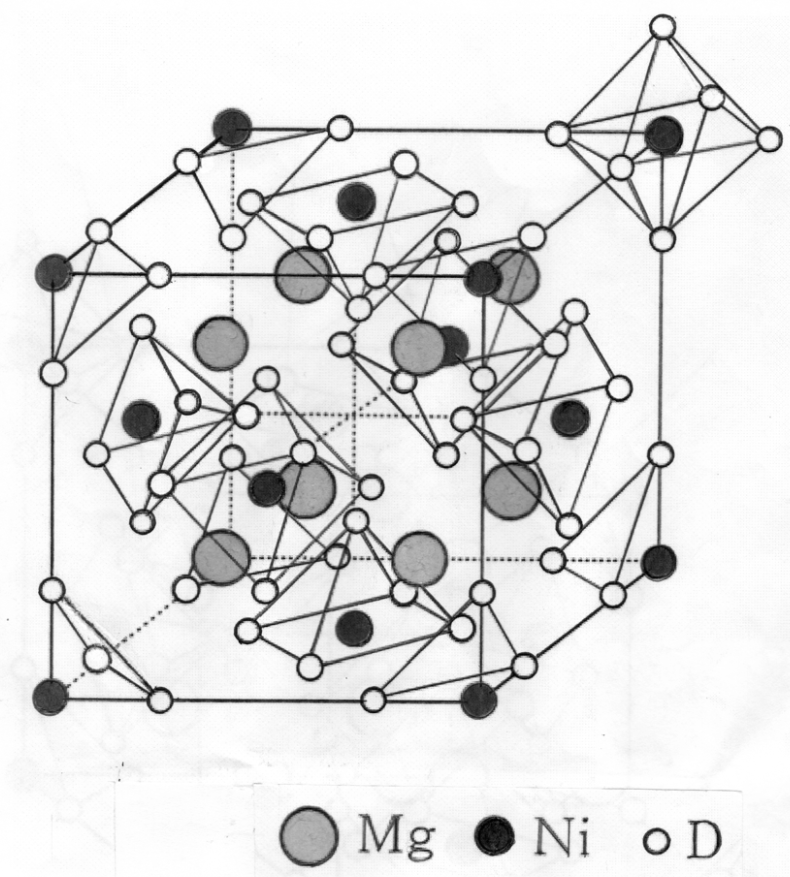}
\end{figure}

\newpage
\noindent
\small
Fig. 1.1 \, Crystal structure of $\mathrm{Mg_2 NiD_4}$ molecules [18, 29, 31]. Hydrogen (H/D) ions are strongly screened and correlated by valence electrons and nearly localized itinerary s-electrons in hybridized states (metallic states) in the molecules immersed in the metallic Li-liquid. These effects are specified by the short range screening length $\mathrm{D_s}$ and the chemical potential change $\mathrm{\Delta G_r}$ in the fusion reaction respectively, and they lead to the enormously enhanced rate of chemonuclear fusion passing through the coherent collapse of hydrogen bonds.

\vspace{0.5cm}
\normalsize
We may charge a test system with $\mathrm{Ni/Mg_2 Ni}$ nanocrystal powder mixed with $\mathrm{LiH/LiAlH_4}$ powder and hydrogen gas. Here, Li atoms play the role of electron donor. With a pressure of 10 bars to the gas, its chemical potential increases by 9kJ/mol=0.09eV/molecule at $T$=773K ($k_{\mathrm{B}}T$=0.067eV) [37] and results in a gain of hydrogen dissolution into the $\mathrm{Ni/Mg_2Ni}$ crystals by a factor of exp(0.09 / 0.067) = 3.8. This implies the gain of fusion rate and hereby power released by a factor of 15. With this treatment we may convert the $\mathrm{Ni/Mg_2 Ni}$ crystals substantially to the hydride $\mathrm{NiH/Mg_2 Ni H_4}$ crystals in which some 10 percent fraction of hydrogen atoms occupy O-sites in the hydride crystals forming 3-pairs as shown in Fig. 1.1. This leads to the trebly coherent collapse of $\mathrm{D_3}$-$\mathrm{D_3}$ bonds with the enhancement of $1.4\times 10^{40}$ $\sim 12\times10^{46}$ at 460K recalling Eq.(28) and Eq.(28').

We begin a test experiment by pre-heating the test system at room temperature. At first, $\mathrm{LiAlH_4}$ molecules dissociate and form $\mathrm{LiH}$ and $\mathrm{AlH_3}$ molecules. The LiH molecules dissociate through physisorption and then chemisorption processes on the surface of $\mathrm{Ni/Mg_2 Ni}$ nano-\\crystals more quickly than hydrogen molecules due to their large dipole moment $p$(LiH)$=2.0\times 10^{-29}(\mathrm{C\cdot m})$ comparable with $p$(LiF)$=2.1\times 10^{-29} (\mathrm{C\cdot m})$ and their weak bond strength $D$(Li-H)$=2.47 (\mathrm{eV})$ compared with $D$(H-H)$=4.52 (\mathrm{eV})$. The dissociated H atoms are then adsorbed by the nano-crystals and transformed into metallic hydrogen leaving Li$^+$ ions and itinerant s-electrons, resulting in the thermody-\\namic activity of liquid together with metallic hydrogen inducing the chemonuclear fusion reactions. The physi- and chemi-sorption processes are also the case where cold nuclear fusion takes place in connection with the electrolysis of $\mathrm{D_2}$O containing LiOD using Pd electrodes.

When the Li ions and itinerant s-electrons moistens the hydride powder if necessary under an activation by Tesla coil corona discharge, the coherent H-H fusion starts up with a large enhancement and heats up the hydride quickly. This produces D-atoms in the hydride together with lattice vacancies which are refilled by H-atoms in the presence of the pressure H$_2$ gas. The D-atoms diffuse then with the diffusion coefficient of ${D_0 \sim 10^{-7} \: m^2/s}$ at a high temperature towards the full clustering at O-sites [38, 39], hereby the trebly coherent $\mathrm{D_3}$-$\mathrm{D_3}$ fusion takes place resulting in the reaction energy released far larger than the H-H fusion as seen in Section 1.13.

Supposedly, the most remarkable feature of H-H fusion in the Ni hydrides is the $\gamma$-ray missing positron annihilation. In this case, most positron annihilation $\gamma$-rays produced in the H-H fusion undergo the electron conversion associated with nuclear charge shake off Auger electron $\textrm{e}^-_A$ emission as
\begin{equation}
\gamma + \textrm{Ni} \longrightarrow \textrm{e}^- + \textrm{e}^-_A + \textrm{Ni}^{2+}.
\end{equation}
In the presence of the liquid activity in the system, this process is enhanced by the factor
\begin{equation}
K = \mathrm{exp} \: \lbrack \frac{-\Delta G_r(\textrm{Ni}^{2+}_{\textrm{liq}})}{k_BT}\rbrack = \mathrm{exp} \: \lbrack \frac{-\Delta G_f(\textrm{Ni}^{2+}_{\textrm{aq}})}{k_BT}\rbrack \approx 2\times10^5,
\end{equation}
$\Delta G_f$ being the formation Gibbs energy, at $T$=460K with $-\Delta G_f(\textrm{Ni}^{2+}_{\textrm{aq}})$ =  0.48 (eV), [16]. Here, $\textrm{Ni}^{2+}_{\textrm{liq}} = \textrm{Ni}^{2+}_{\textrm{aq}}$ (in aqueous solution) has been assumed on the basis of argument in Section 1.2.
Since the value of conversion coefficient is some tenths percent, the enhancement Eq.(33) implies that all annihilation $\gamma$-rays are converted into electrons.
These conversion electrons generate numerous soft x-rays, which have been observed in every successful cold fusion experiment [34, 35, 41]. 

\hyphenchar\font=-1
\section*{1.15 Performance of Chemonuclear Fusion Test System}
\noindent
The theory of chemonuclear fusion predicts the following:
\begin{enumerate}[i)]
\item In a chemonuclear fusion test system charging Li permeated \\$\mathrm{Ni/Mg_2Ni}$ nano-cluster powder and hydrogen gas, we expect the trebly coherent chemonuclear $\mathrm{H_3}$-$\mathrm{H_3}$ and $\mathrm{D_3}$-$\mathrm{D_3}$ fusion reactions but not H-D fusion which generates 5.5MeV $\gamma$-ray.
\item In the system charging ordinary hydrogen gas, contribution from the H-H fusion is only 1.1\% of the total power released. Most power is released in the $\mathrm{D_2}$-$\mathrm{D_2}$ and $\mathrm{D_3}$-$\mathrm{D_3}$ fusion caused by D$_2$ gas which is found by a fraction of about 1/6000 of ordinary hydrogen gas. Even if we take into account the fewer adsorption efficiency of D$_2$ gas by about 8\% compared to H$_2$ gas [40], the contribution from the H-H fusion is still limited to 4.4\%.
\item For the coherent fusion enhancement of about $10^{40}\sim10^{46}$ at $T$=460K and ${n_{D_3}}/{n_D}\approx 10^{-2}$ in the Ni deuteride, we may expect the power output of the $\mathrm{D_3}$-$\mathrm{D_3}$ fusion of $\mathrm{MW \cdot m^{-3}\sim GW \cdot m^{-3}}$ which is far over the solar interior power density.
\item In the coherent $\mathrm{D_n}$-$\mathrm{D_n}$ fusion, the intermediate fusion complexes (quasi-He$_n$) undergo the radiation-less decay and yield He ions of 23.8 MeV kinetic energy. These ions undergo diverse useful successive He induced chemonuclear reactions as explained in Chapters 2, 3 and 4.
\end{enumerate}

\hyphenchar\font=-1
\section*{1.16 Explaining the Mysterious Cold Nuclear Fusion Puzzle}
\noindent
Numerous experiments have been tried unsuccessfully in an attempt to observe the sign of stable self-sustained cold nuclear fusion. Most unsuccessful experiments were concerned with the detection of 
\begin{enumerate}[i)]
\item calorimetrical signals
\item annihilation $\gamma$-rays following the H(p,$\textrm{e}^+$$\nu$)D reaction
\item neutrons produced in the D(d,n) $^3$He reaction
\end{enumerate}
These features are well made clear on the basis of chemonuclear fusion mechanism as follows:
\begin{enumerate}[I)]
\item The instability of cold nuclear fusion was mainly caused by insufficient itinerant electron density in the reaction space such as metal-like hydrides. The essential role of itinerant s-electrons in the chemonuclear fusion hereby the cold nuclear fusion had gone unnoticed for a long time. Without sufficient density of itinerant s-electrons, neither the reduction of melting point of adsorbed metallic hydrogen nor the actualization of its thermo-\\dynamical liquid activity is expected (Sections 1.4-1.5). 
\item It is infeasible to observe the annihilation $\gamma$-rays in the presence of the thermodynamical liquid activity (Section 1.14) due to the enhanced electron conversion process indicated in Eqs.(32) and (33). Similar enhanced electron conversion takes place generally in the metal-like hydride-electron donor mixtures.
\item Unless an appropriate ($\alpha$, n) reaction converter such as a lump of Be is installed in the mixtures, it is unlikely to observe neutrons associated with the cold nuclear fusion. Because the D(d,n)$^3$He reaction is also suppressed relatively in the presence of thermo-\\dynamical liquid activity that enormously enhances the neutron-\\less coherent $\mathrm{D_n}+\mathrm{D_n}\longrightarrow n^{4}$He $(n > 1)$ reaction as argued in Sections 1.8-1.10 and 1.13. 
\item Opposite to the features i), ii) and iii), there are many reports on the spectroscopically and chemically confirmed observation of $^4$He produced in the D-D cold fusion experiments [41]. The reaction $\mathrm{D+D} \rightarrow \mathrm{^4He} + 23.8 \, \mathrm{MeV}$ is, however, inhibited on the basis of two body kinematics. Thus the observation of $^4$He had been one of the prominent puzzle of cold nuclear fusion so far. The observations of $^4$He provided strong evidence for the chemonuclear fusion scheme explained in Sections 1.10 and 1.13.
\end{enumerate}

\hyphenchar\font=-1
\section*{1.17 High Temperature Chemonuclear Reactions in Ni \\Hydride-Li Mixtures}
\noindent
So far we have argued about nuclear fusion at a moderate temperature between $460\sim770\degree$K with an implicit assumption of temperature independent chemical potential. For the system charging Li of some weight percent of Ni or Ni hydride (NiD$_2$ or NiH$_2$), this implicit assumption is no longer adequate. Note that the chemical potential increases with temperature in the presence of enough amount of Li, because the charged Li will provide additional itinerant s-electrons with the increase of temperature.

The electron work function of Li is only 1.4(eV), so the density of $n_e$ of itinerant electrons supplied by Li increases quickly when the temperature goes up, for instance, by a factor of higher orders of magnitude at $1200\degree$K compared to the value at $460\degree$K. While the chemical potential varies proportional to $n_e^{2/3}$ except for extremely high density electrons as argued in Section 1.18. 

The enhancement factor exp[$-\Delta G_\mathrm{r}/T$] does not decrease with the temperature due to the temperature dependence of itinerant electron density and hereby the chemical potential. On the other hand, the chemonuclear reaction rate has a tendency to increase to some extent in this system until $1700\degree$K where the stability of Ni nano cluster powder is no longer guaranteed.

The above argument has been well supported by experiments in Japan and Europe [35, 36, 41].

\hyphenchar\font=-1
\section*{1.18 Thermodynamically Enhanced Pycnonuclear Fusion in White Dwarf Progenitors of Supernovae}
\noindent
We have looked into the mechanism of hydrogen chemonuclear fusion and enhancement in the metal-like hydride-electron donor mixtures. In the presence of itinerant electrons in the metal-like hydride, the melting point of adsorbed metallic hydrogen is reduced, and hence we see the formation of liquid system of hydrogen ions and itinerant electrons. This system reveals the thermodynamical activity of liquid and induces the super-enhanced chemonuclear fusion. However, the enhancement $K=\mathrm{exp}\: \lbrack\frac{-\Delta G_r}{k_BT}\rbrack$ will be reduced drastically at extremely high temperatures. It is a crucial test to investigate whether or not the arguments for the chemonuclear fusion reactions are applicable to the pycnonuclear C-C and C-O fusion in the extremely high temperature condensed plasmas in the white dwarf progenitors of supernovae.
Incidentally we may remark that nuclear fusion rates in dense binary mixtures of carbon and oxygen are essential quantities governing the evolution and ignition in the white dwarf supernova progenitors [14]. First principle calculation of nuclear fusion rates in dense C-O binary ionic mixtures of equi-molar fraction has been done in the fluid phases [42]. The enhancement factors of $10^{24}$ and $10^{30}$ are reported for the C-C and C-O nuclear fusion respectively at the mass density $4\times10^9 \frac{g}{cm^3}$ and the temperature $10^8$K in a state of degree of Fermi degene-\\racy for electrons $\frac{k_BT}{E_F}=2\times10^{-4}$.

Here the author attempts to estimate the chemical potential of elements immersed in high density electron sea by considering their immersion energies [26].

The immersion energies $\Delta E^{hom}$ are formed of terms that can be associated with
\begin{enumerate}[i)]
\item chemical potential effects due to addition of $Z$ electrons on the Fermi level in embedding the atoms of atomic number $Z$ 

and
\item relaxation effects due to improved screening in the metallic\\ environment.
\end{enumerate}
That is, $\Delta E^{hom}$ is written as
\begin{equation}
\Delta E^{hom}(n_e) = Z\bar{\mu}(n_e) + \Delta E_R(n_e),
\end{equation}
where $\bar{\mu}$ is the internal chemical potential of the electron gas and $\Delta E_R$ is the relaxation or rearrangement energy. At high densities the relaxation energy $\Delta E_R$ is only weakly dependent on density, that is, extremely high density limit $\Delta E_R(n_e)\propto n_e^{\frac{1}{6}}$ and will act to compensate the rapid increase $(\propto Zn_e^{\frac{2}{3}})$ in the chemical potential term. 

This fact suggests a dimension of $Z_{eff}$, the effective number of electrons in an atom sensitive to environment as, 
\begin{equation}
Z_{eff}=\frac{\frac{d\Delta E^{hom}}{dn_e}}{\frac{d\bar{\mu}}{dn_e}}
\end{equation}
It is reasonable to evaluate chemical potential $\phi^*(Z)$ of elements of the atomic number Z immersed in the extremely high density electrons as,
\begin{equation}
\phi^*(Z) \sim Z_{eff}\bar{\mu}
\end{equation}
instead of $Z\bar{\mu}$, where
\begin{equation}
\bar{\mu} = 2.6 (\textrm{eV})\: \hspace{0.3cm} \mathrm{at} \hspace{0.3cm}\:  n_e(\mathrm{on\,earth}) = 0.03 \: \mathrm{a.u.}
\end{equation}
provided
\begin{align*}
\mathrm{a.u.} \equiv (4\pi/3)^{-1}\,a_\mathrm{B}^{-3},
\end{align*}
and $Z_{eff}$ is tabulated as [26], 
\begin{align*}
\textrm{C}: 2.97 \: , \textrm{O}: 3.61\:, \textrm{Mg}: 6.39\: , \textrm{Si}: 7.66\,.
\end{align*}
Salient features of enhanced C+C$\rightarrow$Mg and C+O$\rightarrow$Si pycnonuclear fusion in the dense carbon-oxygen matters in the white dwarf (W.D.) are their astronomically huge chemical potential change $\bar{\mu}$ (W.D.), as seen below.
\begin{align*}
- \Delta G_r(\mathrm{C+C})=\phi^*(\mathrm{Mg})-2\phi^*(\mathrm{C})=\big\{Z_{\mathrm{eff}}(\mathrm{Mg})-2Z_{\mathrm{eff}}(\mathrm{C})\big\}\bar{\mu}
\end{align*}
\begin{equation}
=0.45 \bar{\mu},
\end{equation}
\begin{align*}
- \Delta G_r(\mathrm{C+O})=\phi^*(\mathrm{Si})-\phi^*(\mathrm{C})-\phi^*(\mathrm{O}) 
\end{align*}
\begin{equation}
\,\,\,\,\,\,\,\,\,\,\,\,\,\,\,=\big\{Z_{\mathrm{eff}}(\mathrm{Si})-Z_{\mathrm{eff}}(\mathrm{C})-Z_{\mathrm{eff}}(\mathrm{O})\big\}\bar{\mu}=1.08\bar{\mu}\,.
\end{equation}
The values in Eqs.(38) and (39) may be applicable to the rough rate estimation of fusion in the high density plasma in the white dwarf through the correction of $\bar{\mu}$ in Eq.(37).
\begin{equation}
\bar{\mu} (\mathrm{W.D})= \big\{\frac{n_e(\mathrm{W.D.})}{n_e(\mathrm{on \:earth})}\big\}^{\frac{2}{3}} \bar{\mu}\: (\mathrm{on\: earth}) = 10.7\times10^6(\mathrm{eV}). 
\end{equation}
Here, $n_e(\mathrm{W.D.})=2.5\times10^8 \mathrm{a.u.}$ has been assumed [14]. The value in Eq.(40) is evidently over estimate since $\bar{\mu}$ varies proportional to $n_e^{\frac{1}{3}}$ rather than $n_e^{\frac{2}{3}}$ due to the relativistic effect for extremely high density electrons [14]. So, we prefer an intermediate $n_e$ dependence of $\bar{\mu}$ as,
\begin{align*}
\bar{\mu} (\mathrm{W.D})= \big\{\frac{n_e(\mathrm{W.D.})}{n_e(\mathrm{on \:earth})}\big\}^{\frac{5}{9}} \bar{\mu}\: (\mathrm{on\: earth}) = 7.5\times10^5(\mathrm{eV}), \tag{40'}
\end{align*}
and obtain,
\begin{equation}
-\Delta G_r(\mathrm{C+C}) \sim 3.5\times10^5(\mathrm{eV}),
\end{equation}
\begin{equation}
\mathrm{logK}(\mathrm{C+C}) \sim 18\:at\:T=10^8\mathrm{K},
\end{equation}
and
\begin{equation}
-\Delta G_r(\mathrm{C+O}) \sim 8.9\times10^5(\mathrm{eV}),
\end{equation}
\begin{equation}
\mathrm{logK}(\mathrm{C+O}) \sim 44 \:at\:T=10^8\mathrm{K}.
\end{equation}
The first principle calculation has presented [42],
\begin{equation}
\mathrm{logK}(\mathrm{C+C})=23.49\:at\:T=10^8\mathrm{K},
\end{equation}
and
\begin{equation}
\mathrm{logK}(\mathrm{C+O})=29.95 \: at \: T=10^8\mathrm{K}. 
\end{equation}
These values are not very far from the respective values in Eqs.(42) and (44).
As such, super-enhanced chemonuclear fusion in liquid metals and enhanced pycnonuclear fusion in astrophysical condensed plasmas are essentially of the same class.
The only difference is the coherent hydrogen-hydrogen fusion within the correlating hydrogen ion pairs in the same sites of hydrogen adsorbent nano-clusters, as argued in Sections 1.9-1.11.

\hyphenchar\font=-1
\section*{1.19 Concluding Remarks}
\noindent
Over the past century, we have long conceived that nuclear reactions are never influenced by chemical reactions of atoms and molecules surrounding the reactant nuclei. In metallic hydrogen and lithium liquids consisting of collective itinerant s-electrons and nuclei, on the other hand, nuclear reactions are subject to the thermodynamical activity of liquids revealed in linked spontaneous chemical reactions through coupling between the bulk of nuclei and s-electrons.

A typical example is chemonuclear fusion in metallic Li liquids where slow nuclear collisions are unified with atomic collisions and result in the astronomically enhanced nuclear fusion. The s-electron density with respect to the nuclear positions plays the essential role through the coupling between the nuclear transitions and spontaneous atomic and molecular transitions. In those liquids, all extraordinary enhancement mechanisms are strictly subject to the thermodynamic force specified by the chemical potential changes in the reactions.

This conclusion would be an extension of Einstein's unwavering confidence that only thermodynamics is not based on any hypothesis, and will never be overthrown [17].

The chemonuclear reactions will be proved to be a great boon to the human race.

\newpage
\section*{Appendix 1. Thermodynamical Activity of Liquids enhanc-\\ing Chemonuclear Reaction}
\hyphenchar\font=-1
\addcontentsline{toc}{section}{Appendix 1. Thermodynamical Activity of Liquids enhancing Chemonuclear \newline Reaction}
Here, arguments are developed on the mechanism of Chemonuclear Reaction based on the Widom's concept ''thermodynamic activity of liquids'' [13]. 

The configuration integral $Q_N$ for a liquid of $N$ identical particles in a volume $V$ at the temperature $T$ is

\begin{align*}
Q_N = \int_V\cdots \int_V \mathrm{exp} \Bigg( -\frac{W_N}{\mathrm{k_B}T}  \Bigg) \mathrm{d} \tau_1 \cdots \mathrm{d} \tau_N \, , \\
= \int_V\cdots \int_V \mathrm{exp} \Bigg( -\frac{\Psi}{\mathrm{k_B}T}  \Bigg) \mathrm{exp} \Bigg( -\frac{W_{N-1}}{\mathrm{k_B}T}  \Bigg) \mathrm{d} \tau_1 \cdots \mathrm{d} \tau_N \, , \\
= Q_{N-1}V \Bigg\langle \mathrm{exp} \Bigg( -\frac{\Psi}{\mathrm{k_B}T}  \Bigg) \Bigg\rangle ,
\tag{A1}
\end{align*}
where $\mathrm{d} \tau_1 \cdots \mathrm{d} \tau_2$ is the element of volume in the configuration space of the $N$ molecules, $W_N$ is the potential energy of interaction of the $N$ particles such as atoms or molecules as a function of their positions in the volume $V$, and $\Psi$ is the potential energy of interaction of one molecule with the remaining $N-1$ molecules as a function of the positions of all $N$ of them. The mean value symbolized by $\langle \, \rangle$ is a canonical average.

Let $z$ be the thermodynamic activity of the liquid molecules; it is defined so at so become asymptotic to the number density $N/V\equiv n$ in the limit $n\rightarrow0$. Then
\begin{align*}
z = \frac{NQ_{N-1}}{Q_N}, \\
\mathrm{so \,from\, Eq. \,(A1)}, \\
n/z = \Bigg\langle \mathrm{exp} \Bigg( -\frac{\Psi}{\mathrm{k_B}T}  \Bigg) \Bigg\rangle .
\tag{A2}
\end{align*}
In a mixture of several kinds of molecules, if $n$ and $z$ are the density and activity of one species, and $\Psi$ the interaction energy of a molecule of that species with the rest of the liquid, then eq. (A2) still holds. 

If the liquid consists generally of molecules of one kind (solvent), and $\Psi$ is the energy of the interaction of molecules of another kind (solute) with a solvent liquid, then $\langle$exp$(-\Psi/\mathrm{k_B}T)\rangle$ is basically the reciprocal of the Henry's law constant, and this term corresponds to the thermodynamical enhancement factor exp$(-\Delta G/\mathrm{k_B}T)$ in Eq.(16). In what follows, however, the concern is entirely with pure liquids.

In Eq.(A2) a solute molecule interacts with all of the molecules of the solvent liquid instead of a two-body interaction with an isolated liquid molecule. The thermodynamic activity of liquids is thus found to be a bulk/collective feature caused by the thermodynamic force in the liquid and the derivation of Eq.(A2) is a consequence of a major proposition that the solvent liquid consists of interacting molecules or particles at thermal equilibrium.

The thermodynamic activity of metallic liquids is closely related to the cohesion of metals produced by collective itinerant electrons. In fact, the thermodynamic enhancement factor exp$(-\Delta G / \mathrm{k_B}T)$ of the p+Li and d+Li reactions has been derived from the bond energy of liquid Li metal in ref. 6. Each atom in a metal covalent bond shares an electron with its nearest neighbors, but the number of orbitals for bond formation exceeds the number of electron pairs available to fill them. As a result, covalent bonds resonate among the available interatomic positions. This resonance extends throughout the entire structure, thereby producing greatly enhanced stability. Another bulk feature of metals is also seen in the difference of electric conductivity between metals (e.g. Ag) and typical insulators by a factor up to some 20 orders of magnitude.

Note that certain bulk features of matters are due to macroscopic quantum effects caused by coherent dynamics of itinerant valence electrons of which de Broglie wave lengths stretches over a bond length of some tens metallic atoms for instance Li atoms at melting point. These effects can be treated in the scheme of statistical thermo-\\dynamics. Such features may be expected in certain nuclear reactions through atomic fusion, that is, united atoms formation enhanced by the thermodynamic force in metallic liquids [6].

When implanted ions are travelling in a metallic liquid, atoms of the liquid surrounding the ions tend towards formation of a meta stable phase such as an alloy or compound at the minimum Gibbs energy point. However, this Gibbs energy is not the minimum point any longer for the buffer energy ($ \leq 25$ keV/amu) ions. The buffer energy ions penetrate the Coulomb barrier of liquid atoms to form a new reaction intermediate, that is, united atoms at their turning points. Physical characteristics such as the Gibbs energy and density of united atoms are expected to be almost the same with those of product atoms of nuclear fusion [6] implying the possibility of nuclear fusion induced by united atoms.

For the $^6$Li(p.$\alpha )^3$He and $^7$Li(p.$\alpha )^4$He reactions via the formation of united atoms $\overline{\mathrm{LiH}}$ and excited nuclei $^{7,8}$Be$^*$ in a metallic Li liquid,

\begin{align*}
^{6,7}\mathrm{Li+\,^1H}\rightarrow \overline{\mathrm{LiH}} \rightarrow ^{7,8}\mathrm{Be}^*,
\tag{A3}
\end{align*}
the reaction Gibbs energy has been estimated using the bond energy of metallic Li liquid to be about $\Delta G= -1.25$eV for a reacting atom pair in [6] which agrees with the value $\Delta G_r= -1.35$eV in Table 2.1. This, in turn, results in an enhancement of fusion reaction rate by a factor
\begin{align*}
\mathrm{exp} \Bigg[ -\frac{\Delta G}{\mathrm{k_B}T} \Bigg] = 4.6\times10^{13},
\tag{A4}
\end{align*}
just above the melting point of metallic Li, $T=460$K [6]. 

\newpage

\section*{Appendix 2. Quenching Effects on Chemonuclear Reaction}
\hyphenchar\font=-1
\addcontentsline{toc}{section}{Appendix 2. Quenching Effects on Chemonuclear Reaction}
The enhancement argued in Appendix 1, however, has never been observed in nuclear reactions due to possible quenching effects.

Consider a case where hydrogen ions of buffer energy $E_\mathrm{Lab} \leq 25$ KeV/amu are implanted into a metallic Li liquid around its melting point. The ions quickly become neutralized in the metal. On traversing the metal, however, the neutralized particles ionize again. The depth, in the metal at which the equilibrium charge state is achieved, is related to the atomic density or electron density of the metal and also the average electron capture and loss cross sections. This depth corresponds to several tens of monolayers of metal atoms.

Certain hydrogen ions undergo collisions with nuclei of metallic Li atoms, because the nuclear stopping is marked for buffer energy ions [6]. During the collisions, an ion transfers considerable amounts of recoil momentum and kinetic energy to an interacting Li atom. It is unlikely that the recoil Li atom will be at thermal equilibrium with the rest of the liquid Li atoms. In this case, the reaction rate would deviate from that of the Arrhenius' equation, thereby quenching the thermodynamic enhancement exp$(-\Delta G/\mathrm{k}T) >1$.

The colliding Li atom receives a recoil energy of about $E_\mathrm{R}=1\sim2$keV from the buffer energy ion without its electronic excitation as argued in [1]. The Li atom is thus not hot and still keeps a correlation with the rest of Li atoms to some extent. Recalling the arguments in Appendix 1, the ensemble of metallic atoms at thermal equilibrium is regarded as a thermodynamic resonator which may be specified by zero resonance energy and the half width of k$_\mathrm{B}T$. The recoil process can be treated as a problem of resonance scattering of particles by the resonator and the quenching or correlation factor $\rho(E_\mathrm{R},T)$ of the enhancement is expressed in a simple resonance formula. One can derive the formula as a correlation function of the potential activity of the recoil atom defined by exp$(-\Psi/E_\mathrm{R})$ and the thermodynamic activity of the rest atoms $\langle$exp$(-\Psi/\mathrm{k_B}T)\rangle$ derived by the equation

\begin{align*}
\rho(E_\mathrm{R},T) = \mathrm{Re}\Bigg[ \int_{0}^{\infty} \Bigg\langle \mathrm{exp} \Bigg( -\frac{\mathrm{i}\Psi}{\mathrm{k_B}T}  \Bigg) \Bigg\rangle \mathrm{exp} \Bigg( -\frac{\Psi}{E_\mathrm{R}} \Bigg) \mathrm{d} \Bigg( \frac{\Psi}{E_\mathrm{R}} \Bigg) \Bigg] \\= \frac{(\mathrm{k_B}T)^2}{E_\mathrm{R}\,^2+(\mathrm{k_B}T)^2}\, .
\tag{A5}
\end{align*}
This results in a quenching of the reaction rate Eq.(1) as,

\begin{align*}
k_{12}(T)= A_{12} \cdot \rho(E_\mathrm{R},T)\mathrm{exp} \Bigg( -\frac{\Delta G}{\mathrm{k_B}T}  \Bigg)\,.
\tag{A6}
\end{align*}

When a beam proton is implanted with an acceleration energy of 10 keV into a Li liquid at its melting point 454 K, the proton transfers a recoil energy of 1.25 keV to a metallic $^7$Li atom during their interaction, and the correlation factor in Eq.(A5) would be

\begin{align*}
\rho(E_\mathrm{R},T) \sim 5.7\times10^{-10}\,.
\tag{A7}
\end{align*}
which reduces the effective enhancement up to $\rho(E_\mathrm{R},T)\cdot(-\Delta G/\mathrm{k_B}T) \sim 2.6\times10^4$ recalling the unquenched value $4.6\times10^{13}$ in Eq.(A4).

The reduced value of effective enhancement agrees well with the Uppsala and Tokyo data of preliminary experiments on the buffer energy $^7$L(p.$\alpha)^4$He reaction [7, 9]. In both experiments, proton beams extracted from ion sources entered target chambers equipped with Si detectors and were injected vertically on metallic Li liquid surfaces. The reaction product $\alpha$-particles of 8.6 MeV were observed with the detectors positioned at the angles $\theta_\mathrm{Lab}$115$\degree$ (Uppsala) and 135$\degree$ (Tokyo).

\newpage

\section*{References}
\begin{enumerate}
\small
\item S. Ichimaru and H. Kitamura, Phys. Plasma \textbf{6} (1999) 2649.
\item A.G.W. Cameron, Astrophys. J. \textbf{130} (1959) 916.
\item B. Jancovici, J. Stat. Phys. \textbf{17} (1977) 357.
\item S. Ichimaru and H. Kitamura, Publ. Astron. Soc. Jpn. \textbf{48} (1996) 613. \\S. Ichimaru: Rev. Mod. Phys. \textbf{65} (1993) 255.
\item S. Ichimaru, Phys. Lett. \textbf{A266} (2000) 167.
\item H. Ikegami, Jpn. J. Appl. Phys. \textbf{40} (2001) 6092. See also No. 1 and No. 2 papers in [7].
\item H. Ikegami and R. Pettersson, “Evidence of Enhanced Nonthermal Nuclear Fusionl” Bull. Inst. Chem. (Uppsala Univ. Uppsala, September 2002) No. 3.
\item H. Ikegami, “Ultradense Nuclear Fusion in Metallic Lithium Liquid”, Rev. Ed. of Swedish Energy Agency Document ER2006: \textbf{42} (Sakaguchi, Tokyo/Sakura, 2007) \textbf{3}-1 - \textbf{3}-30.
\item H. Ikegami and T. Watanabe, ibid. \textbf{4}-1 - \textbf{4}-21.
\item H. Ikegami, ibid. \textbf{1}-1 - \textbf{1}-22. The term “chemonuclear fusion” was coined by S. Kullander.
\item H. Ikegami and T. Watanabe, ibid. \textbf{2}-1 - \textbf{2}-12.
\item H. Ikegami et al, J. Phys. Soc. Jpn. Suppl. \textbf{24} (1968) 167.
\item B.Widom, J. Chem. Phys. \textbf{39} (1963) 2808.
\item S. Ichimaru, Rev. Mod. Phys. \textbf{65} (1993) 255.
\item H. Ikegami, R. Pettersson and L. Einarsson, Prog. Theo. Phys. Suppl. \textbf{154} (2004) 251.
\item G. Aylward and T. Findlay: SI Chemical Data ed.3 (Wiley, Brisbane, 1998).
\item D. Kondepudi and I. Prigogine: Modern Thermodynamics (Wiley, Chichester, 1998).
\item Y. Fukai: The Metal Hydrogen System (Springer, 1993), Chap. 6.
\item E. Wigner and F. Seitz, Phys. Rev. \textbf{43}  (1933) 804.
\item W.A. Fowler, G.R. Caughlan and B.A. Zimmerman, Ann. Rev. Astron. Astrophys. \textbf{5} (1967) 523.
\item A. Krauss, H.W. Becker, H.P.  Trautvetter, O. Rolfs and K. Brand, Nucl. Phys. \textbf{A465} (1987) 150.
\item A.R. Miedema, J. Less-Common Metals \textbf{32} (1973) 117.
\item E. Rutherford, Phil. Mag. \textbf{21} (1911) 212; Ibid \textbf{21} (1911) 699.
\item N. Bohr, Phil. Mag. \textbf{25} (1913) 10.
\item R.A. Weller, Ion-Solid Interaction, S.P. Parker Ed.: McGraw-Hill, New York, 1992; McGraw-Hill Encyclopedia of Science and Technology, 7th ed., Vol. 9, p 384.
\item M.J. Puska, R.M. Nieminen and M. Manninen, Phys. Rev.  \textbf{B24} (1981) 3037.
\item H.H. Andersen and H.L. Bay, J. Appl. Phys. \textbf{45} (1974) 953.
\item H. Ikegami, Int. J. Quant. Chem. \textbf{83} (1999) 71.
\item Y. Fukai and H. Sugimoto, Adv. in Phys. \textbf{34} (1985) 263.
\item P. Nordlander, J.K. Nørskov and F. Besenbacher, J. Phys. F \textbf{16} (1986) 1161.
\item G. Sandrock, S. Suda and L. Schlapbach, Hydrogen in Intermetallic Compounds II, L. Schlapbach ed. (Springer, 1992), Chap. 5.
\item R. Boom, F.R. de Boer and A.R. Miedema, J. Less-Common Metals \textbf{46} (1976) 271.
\item E.N. Kaufmann, R. Vianden, J.R. Chelikowsky and J.C. Phillips, Phys. Rev. Lett. \textbf{39} (1977) 1671.
\item H. Ikegami, The Nature of the Chemonuclear Transitions, The Svedberg Laboratory, Uppsala University. Uppsala. 2012. No. 3.
\item A. Rossi, Communication to H. Ikegami, S. Kullander and R. Pet-\\tersson 2012.
\item H. Ess\'en and S. Kullander, Experimental test of a mini-Rossi device at the Leonardo Corp., Bologna 29 March 2011. https://www.lenr-canr.org/acrobat/EssenHexperiment.pdf
\item H. Sugimoto and Y. Fukai, Acta Metall. \textbf{40} (1992) 2327.
\item J.  Völkl and Alefeld: Hydrogen in Metals I (Springer, 1978), Chap. 12.
\item H. Wipf: Hydrogen in Metals III (Springer, 1997), Chap. 3.
\item H. Conrad, G. Ertl and E.E. Latta, Surf. Sci., \textbf{41} (1974) 435.
\item There are many reports on the $^4$He and soft x-rays observation in Europe, Japan and US, for instance, A.De Nnine et al, Experimental evidence of $^4$He production in a cold fusion experiment, RT 2002/41/FUS distributed by ENEA Centro Ricerche Frascati, C.P. 65-00044 Frascati Rome.
\item S. Ogata, H. Iyetomi and S. Ichimaru, Astrophys. J. \textbf{372} (1991) 259.
\item H. Ikegami, Phys. Rev. Lett. \textbf{64} (1990) 1737, ibid. 2593. Coherent charged particle beam generation through CMC was reconfirmed by K. Shimoda and J. Kondo.

\end{enumerate}
\newpage
\LARGE
\noindent
\begin{center}
\textsc{How I Discovered the Chemonuclear Reactions}
\end{center}
\Large
\hyphenchar\font=-1
\section*{}
\normalsize
\noindent 
In the famous Bohr-Einstein controversy in 1927, Einstein warned the Copenhagen school fellows who believed in the Bohr-Heisenberg uncertainty interpretation with “\textit{God does not play dice with the universe}”. After that, Einstein broke away from his Copenhagen school friends and devoted himself to developing new fields from his own standpoint. 

However, what does “God” really mean here?

God is the absolute being. Given the circumstances, Einstein's remark seemed to be a parting shot from a loser in a scientific debate. Quite a few textbooks on quantum mechanics made comments on this matter as if Bohr-Heisenberg won the day. Nonetheless, is there any truth in the comments?

Einstein made another famous remark -- “\textit{Only thermodynamics will never be overthrown}”. To give a faithful account of Einstein's idea, this author quotes his full speech from the preface of \textit{Modern Thermodynamics} [D. Kondepudi and I. Prigogine, 1998]:

``A theory is more impressive the greater the simplicity of its premises is, the more different kinds of things it relates, and the more extended area of applicability. Therefore the deep impression which classical thermodynamics made upon me. It is the only physical theory of universal content concerning which I am convinced that within the framework of the applicability of its basic concepts, it will never be overthrown." 

Insightful readers are aware that the concept of quantum was developed from Planck's thoroughly thermodynamical considerations, and Einstein built his theory of relativity on the classic model of thermodynamics.

Nowadays the Big Bang is universally evident and being treated as the most spectacular spontaneous phenomenon. Based on thermo-\\dynamical considerations, such a phenomenon doubtless points to the actual existence of absolutely irreversible great actions that caused the Big Bang and such actions have governed the whole macrocosm in the evolution of stars and life. Einstein designated none but the absolutely irreversible great actions God.

Einstein was already predicting the \textit{parity non-conservation} when he established the theory of relativity in 1905 (i.e., five decades before the discovery of parity non-conservation in 1956). With the absolutely irreversible actions, time is no longer symmetric and the past and the future are incommutable. Thus the symmetricity of space may also be broken due to the space-time covariant characteristics based on the theory of relativity.

Throughout his life, Bohr had certainly felt guilty about what he had done by following the outgoing W. Heisenberg's argument while neglecting the advice of Rutherford and other senior fellows (Bohr was once a postdoctoral fellow under Rutherford). Interested readers may well find such remorseful accounts from Bohr in his late life.

Supposedly, spontaneous chemical reaction (e.g., burning in the air and ionic reactions in the liquids) is the most popular irreversible phenomena. Some kinds of ionic liquids undergo rapid reactions with the astronomically large enhancement factor of almost $10^{50}$ in the presence of \textit{thermodynamical liquid activity} (i.e., Widome's general concept) as detailed in this book. The present author argues that these remarkable features are induced by the collective dynamics of bulk of atomic electrons revealing the thermodynamical liquid activity.

\textit{In the nuclear reactions in metallic systems with the built-in thermodynamical liquid activity, the bulk of itinerant s-electrons are undergoing contact interaction with reactant nuclei and induce contagiously the thermodynamical liquid activity (i.e., macroscopical scale correlations) among the bulk of nuclei under the irreversible action of Nature towards the chemical potential minimum} of the systems. With this “liquid activity contagion”, united atomic and nuclear reactions, namely the chemonuclear reactions, are induced. The chemonuclear reactions are very likely to be enhanced with the astronomical figures associated with the big chemical potential drop in the reactions. 

This book presents evidence for the chemonuclear reactions, i.e., the observation of coherent chemonuclear fusion, cascade enhanced nuclear reactions, chemonuclear fission and stimulated emission of coherently line-up $\alpha$-cluster. \textit{The current evidence will press us for a fundamental correction of the century-long widely accepted recognition that nuclear reactions are independent from any atomic states of reactants. On the contrary, nuclear reactions in the macroscopic scale are subjected to the chemical potentials} regulating the linked atomic or chemical reactions in the scheme of chemonuclear reaction.

Considering the universal aspect of this argument, we may expect a striking thermodynamical enhancement mechanism in various kinds of systems, irrespective of the kinds of reactant particles or the nature of microscopic inter-particle interactions.

Similar to those super-enhanced nuclear reaction in condensed plasmas in stars (e.g., extraordinarily enhanced pycnonuclear fusion in supernova progenitors), the chemonuclear reactions will be the most enhanced nuclear reaction on earth.

For decades, major countries have been eager to push themselves forward with “hydrogen-hydrogen nuclear fusion in gas plasmas” in the belief that “A Sun on Earth” is the hope for the future. As a result, a great many facilities have been built thanks to the enormous amount of budget for this purpose. Nonetheless, we have to remember that the \textit{released power density of realized chemonuclear $\mathrm{D}_n-\mathrm{D}_n$ fusion is over the factor of one billion compared to the solar interior power density which is the final goal of gas plasma nuclear fusion scientists.}

This book highlights the union of Einstein's warning and the con-\\ception of chemonuclear reaction induced by thermodynamical liquid activity revealed contagiously among the bulk of itinerant s-electrons and reactant nuclei/hadrons.

From the late 20th century to the early 21st century, remarkable nuclear reaction rate enhancements in the condensed plasmas in stars were predicted based on the quantum statistical treatments by astro-\\physicists [1]-[5]. For example, the enhancement by a factor of some 30 orders magnitude were expected in the metallic hydrogen liquid plasmas in a white-dwarf progenitor of a supernova. This kind of enhancement can be attributed to the thermodynamical liquid activity [6] revealed in the dense itinerant s-electrons and reactant nuclei in the metallic liquids.

In 2001 another astronomically enhanced $^7$Li(D, n)$^8$Be$\longrightarrow 2\cdot ^4$He reaction rate in the metallic Li liquid was pointed out based on the microscopic consideration on the slow ion collision process by H. Ikegami [6]. H. Ikegami predicted the rate enhancement $K(\mathrm{Be})=\mathrm{exp}[-\Delta G_\mathrm{r} \mathrm{(Be)}/k_\mathrm{B}\mathrm{T}]$ specified by chemical potential drop $-\Delta G_\mathrm{r}$(Be) in the intermediate united atom (quasi Be-atom) formation in the reaction.
It may be said in passing that the reality of quasi-atom was already confirmed in the late 20th century in the heavy ion colliding experiments.

The predicted enormous rate enhancements was established by detecting $\alpha$-particles produced under slow D-ions implantation on a Li-liquid target by H. Ikegami and R. Pettersson in Uppsala [8]. Since then the new reaction has been called chemonuclear reaction as per S. Kullander's coined word.

In this new scheme of reaction in metallic Li-liquid, even doubly intensified enhancement $K(\mathrm{Be}_2)=K^2(\mathrm{Be})=\mathrm{exp}[-2\cdot\Delta G_\mathrm{r} \mathrm{(Be)}/k_\mathrm{B}\mathrm{T}]$ was expected with the coherent formation of intermediate quasi-Be$_2$ molecule coherently under the slow D$_2$-ion implantation [9].

Hence the coherent collisions were known in the 1970's in the coherent sputtering phenomena of molecular ions [10]. This coherent chemonuclear reaction was proven by the author through a comparison experiment of atomic and molecular ion implantation on the same Li-liquid target in collaboration with T. Watanabe in Tokyo/Sakura [9].

He had enforced the above experiments in spite of the disturbance caused by the deterioration of extremely reactive metallic Li liquids targets. When investigating the function of Ni nanocrystal powder as a stabilizer for the Li liquid, he found the specially significant characteristics of the system of hydrogen adsorbent Ni nanocrystal powder and Li and/or Li hydride mixture, more generally, hydrogen adsorbent nanocrystal powder and electron donor mixture. In the early 1970s, metal scientists pointed out the possible formation of metallic hydrogen in the adsorbents through their systematic experiments with alloys [11].

Those metal scientists found that in the adsorbents, hydrogen or Li hydride molecules dissociate at the surfaces of adsorbent Pd or Ni nano-crystals. They are transmuted into metallic hydrogen and Li releasing itinerant s-electrons respectively. According to the quantum statistical treatments, the electrons are expected to induce the lowering melting point of adsorbed metallic hydrogen [1]. Under the existence of the dense itinerant s-electrons, the system reveals thermodynamical liquid activity among the bulk of reactant nuclei that come to undergo astronomically enhanced nuclear reactions. 

As a result, astronomically enhanced coherent D$_2$-D$_2$ fusion and D$_3$-D$_3$ fusion take place, thereby producing energetic (23.8 MeV) He ions. These He ions provoke diverse cascade enhanced chemonuclear reactions among the reactants charged in the system.

In sum, the system manifests the nature of the most active reactor that has ever been built, i.e., hybrid fusion/fission reactors, radioactive waste vanishing reactions, a variety of element synthesis systems.

Various pieces of evidence support the chemonuclear reactions. The chemonuclear reactions are key to our revision of the traditional belief that nuclear reactions are independent from any atomic states of reactants. More importantly, the chemonuclear reactions taking place on a macroscopic scale are subject to chemical potentials concerning the metallurgy, chemistry and physics of nuclei and hadrons. There is room for chemonuclear research on cosmology.

Furthermore, it might be possible for thermodynamical treatments in chemonuclear reactions to be applied to social science problems because the above treatments are expected to be applicable to various kinds of reactant systems irrespective of their kinds and the nature of interaction within the individual reactants in the macrocosmos. This implies that the thermodynamical treatments presented in this book are applicable to social science problems through introducing proper social chemical potential such as social or global unrest.

Thoughtful readers would find that the arguments presented in this book defer to Einstein's remark that only thermodynamics will never be overthrown.

The author has spent two decades developing the concept of the chemonuclear reactions, i.e.,  macroscopically unified chemical and nuclear reactions. Initially, the author surmised that the enormously enhanced pycno-nuclear fusion in the condesed plasmas in the white dwarf progenitor of supernovae might be realized on earth too, in concrete terms, in the itinerant s-electron-rich liquid metallic medium. He noted the possible occurrence of new enhanced pycno-nuclear fusion in his \textit{Jpn. J. Appl. Phys.} article in 2001.

The first evidence came from Uppsala University. The observed fusion rate enhancement in the slow D/D$_2$ ion implantation on metallic Li liquid target was around $10^{10}$. This result was obtained with the support of Uppsala University and the Swedish Energy Agency. This reaction was named chemonuclear fusion or chemonuclear reactions in general. 

The $^7$Li(D, n)$^8$Be$\longrightarrow 2\cdot^4$He chemonuclear reaction, together with the 2Li(D$_2$, 2n)$2\cdot^8$Be$\longrightarrow 4 \cdot ^4$He coherent chemonuclear reaction, was confirmed in Tokyo/Sakura where  the $^7$Li($^7$Li, 2n)$^{12}$C chemonuclear reaction was also observed. These results seemed to suggest that the enhanced pycno-nuclear reactions took place in the metallic Li liquids and in the condensed plasmas in the supernova progenitor as well. 

The findings were submitted to \textit{Jpn. J. Appl. Phys.}. The editor replied informing that the paper dealt with essential and fundamental issues so it should be submitted to \textit{J. Phys. Soc. Jpn.} instead. The editor of that journal, however, unthinkingly suggested rewriting the paper into two parts dealing with chemonuclear reaction theory and related experiments separately. The author then submitted the paper to \textit{Nature} and got a reply saying that the paper dealt with too many subjects giving rise to quite a few new concepts and it was far over the scale of a journal article. Thus he recommended publishing the paper as a book or a collection of papers.

In addition, the author found that some hydrogen adsorbents such as Ni and Pd nanocrystal powders were the most powerful stabilizer for the metallic Li liquids which were extremely unstable even in high vacuum. He began to wonder if the mixtures of metal-like hydride of these metals and itinerant s-electron donors such as Li and CaO would realize the most ideal systems for the wide area of chemonuclear reactions as seen in this text. This surmise was confirmed in Bologna through collaboration experiments of Uppsala and Bologna groups.

This book transcribes the author's decades-long research into the chemonuclear reactions. Hopefully, the contents may stimulate the readers to explore that newly developed promising area.

\newpage
\section*{References}
\small
\begin{enumerate}
\item S. Ichimaru, Rev. Mod. Phys. \textbf{65} (1993) 255.
\item B. Jancovici, J. Stat. Phys. \textbf{17} (1977) 357.
\item S. Ichimaru and H. Kitamura, Publ. Astron. Soc. Jpn. \textbf{48} (1996) 613.
\item S. Ichimaru and H. Kitamura, Phys. Plasma \textbf{6} (1999) 2649.
\item S. Ichimaru, Phys. Lett. \textbf{A266} (2000) 167.
\item B. Widom, J. Chem. Phys. \textbf{39} (1963) 2808.
\item H. Ikegami, Jpn. J. Appl. Phys. \textbf{40} (2001) 6092.
\item H. Ikegami and R. Pettersson, \textit{Evidence of Enhanced Nonthermal Nuclear Fusion} Bull. Inst. Chem. (Uppsala Univ., Uppsala September 2002) No. 3.
\item H. Ikegami, Ultradense Nucl. Fusion in Metallic Li Liquid No. 4 in Revised Edition of Swedish Energy Agency ER 2006-42.
\item H. H. Andersen and H. L. Bay, J. Appl. Phys. \textbf{45} (1974) 953.
\item A. R. Miedema, J. Less-Common Metals \textbf{32} (1973) 117.
\end{enumerate}
\end{document}